\documentclass[a4paper,fleqn,usenatbib]{mnras}

\usepackage{newtxtext,newtxmath}

\usepackage[T1]{fontenc}
\usepackage{ae,aecompl}

\usepackage{graphicx}	
\usepackage{amsmath}	
\usepackage{amssymb}	
\usepackage{color}
\usepackage{csquotes}

\definecolor{RED}{RGB}{255,0,0}
\newcommand{\corr}{}
\definecolor{blue}{RGB}{0,0,255}

\title[Generic Grid of Transmission Spectra]{Fully scalable forward model grid of exoplanet transmission spectra}

\author[J. M. Goyal et al.]{Jayesh. M. Goyal $^{1}$\thanks{E-mail: jgoyal@astro.ex.ac.uk},
Hannah. R. Wakeford$^{2}$,
Nathan. J. Mayne$^{1}$,
\newauthor Nikole. K. Lewis$^{3}$,
Benjamin. Drummond$^{1}$,
David. K. Sing$^{1,4}$
\\
$^{1}$Astrophysics Group, Physics Building, Stocker Road, University of Exeter, Devon EX4 4QL, UK\\
$^{2}$Space Telescope Science Institute, 3700 San Martin Drive, Baltimore, MD 21218, USA\\
$^{3}$Department of Astronomy and Carl Sagan Institute, Cornell University, 122 Sciences Drive, Ithaca, NY, 14853, USA\\
$^{4}$Department of Earth and Planetary Sciences, Johns Hopkins University, Baltimore, MD, USA
}

\date{Accepted October 29th 2018. Received October 15th 2018; in original form July 31st 2018}

\pubyear{2018}

\begin{document}
\label{firstpage}
\pagerange{\pageref{firstpage}--\pageref{lastpage}}
\maketitle

\begin{abstract}

\corr{Simulated exoplanet transmission spectra are critical for planning and interpretation of observations and to explore the sensitivity of spectral features to atmospheric thermochemical processes. We present a publicly available generic model grid of planetary transmission spectra, scalable to a wide range of H$_2$/He dominated atmospheres.} The grid is computed using the 1D/2D atmosphere model \texttt{ATMO} for two different chemical scenarios, first considering local condensation only, secondly considering global condensation and removal of species from the atmospheric column (rainout). The entire grid consists of \corr{56,320} model simulations across 22 equilibrium temperatures (400 - 2600\,K), four planetary gravities (5 - 50 ms$^{-2}$), five atmospheric metallicities (1x - 200x), \corr{four C/O ratios (0.35 - 1.0)}, four scattering haze parameters, four uniform cloud parameters, and two chemical scenarios. We derive scaling equations which can be used with this grid, for a wide range of planet-star \corr{combinations}. We validate this grid by comparing it with other model transmission spectra available in the literature. \corr{We highlight some of the important findings, such as the rise of SO$_2$ features at 100x solar metallicity, differences in spectral features at high C/O ratios between two condensation approaches, the importance of VO features without TiO to constrain the limb temperature and features of TiO/VO both, to constrain the condensation processes.} Finally, this generic grid can be used to plan future observations using the HST, VLT, JWST and various other telescopes. The fine variation of parameters in the grid also allows it to be incorporated in a retrieval framework, with various machine learning techniques.
\end{abstract}

\begin{keywords}
planets and satellites: atmospheres--planets and satellites: composition--planets and satellites: gaseous planets--techniques: spectroscopic
\end{keywords}

\section{Introduction}
\label{section:intro}
\corr{Planning and interpretation of exoplanet atmospheric characterization observations necessarily rely on theoretical spectra generated from atmospheric models. There currently exist several exoplanet atmospheric forward model grids and simulators providing publicly available theoretical transmisison spectra \citep[e.g.][]{Fortney2010, Molliere2016, Kempton2017, Goyal2018} all of which make specific choices about the atmospheric physics, chemistry, radiative transfer and spectroscopic line lists incorporated into their models. However, inter-comparisons between these modeling frameworks still prove difficult. Many of the exoplanet spectral databases produced to date cover very specific non-overlapping parts of exoplanet atmospheric phase space. Often choices made by specific teams concerning the underlying spectroscopic line lists and physical processes like condensation and rainout are not clearly outlined, which can lead to disagreements and confusion when the models are applied by scientists outside the team. Furthermore, the sensitivity of these theoretical spectra to these specific physics choices are often not explored. Publicly accessible databases that provide representative theoretical spectra for exoplanet atmospheres spanning a broad range of atmospheric properties and physical assumptions are necessary to determine the validity of our understanding of these distant worlds to our theoretical constructs. These grids also provide a straight forward way to test spectral sensitivity both within a given modeling framework and across modeling frameworks.}

Transmission spectra observations of exoplanet atmospheres have been increasing in quality and resolution since the commissioning of the Wide Field Camera 3 (WFC3) instrument on board the Hubble Space Telescope (HST) (e.g. \citealt{Deming2013,Kreidberg2014,Wakeford2016,Sing2016,Evans2016,Evans2017,Wakeford2018}) and FORS2 on the Very Large Telescope (VLT) \citep{Nikolov2018}. This increase in data fidelity has also motivated development of  grids of models covering fine variations of model parameters, especially those that alter the chemistry most significantly. 



Robust line-lists are at the core of any spectral generation tool. The HITRAN (High Resolution TRANsmission) database \citep{Rothman2013} has been the source of spectral line-lists for most of the previous models. This database is established at a reference temperature of 296\,K \citep{Rothman2010}, with HITEMP \citep{Rothman2010} its high temperature version available only for certain molecules. Some hot Jupiter exoplanet atmospheres can reach temperatures as high as 3000\,K, where HITRAN line-lists can underestimate absorption of radiation by several orders of magnitude.  However, more accurate spectral line-lists for hot atmospheres have been developed by the EXOMOL project \citep{Tennyson2016}, which has also motivated development of updated more accurate grids of models. 

 
                 

\begin{figure*}
	\includegraphics[width=\textwidth]{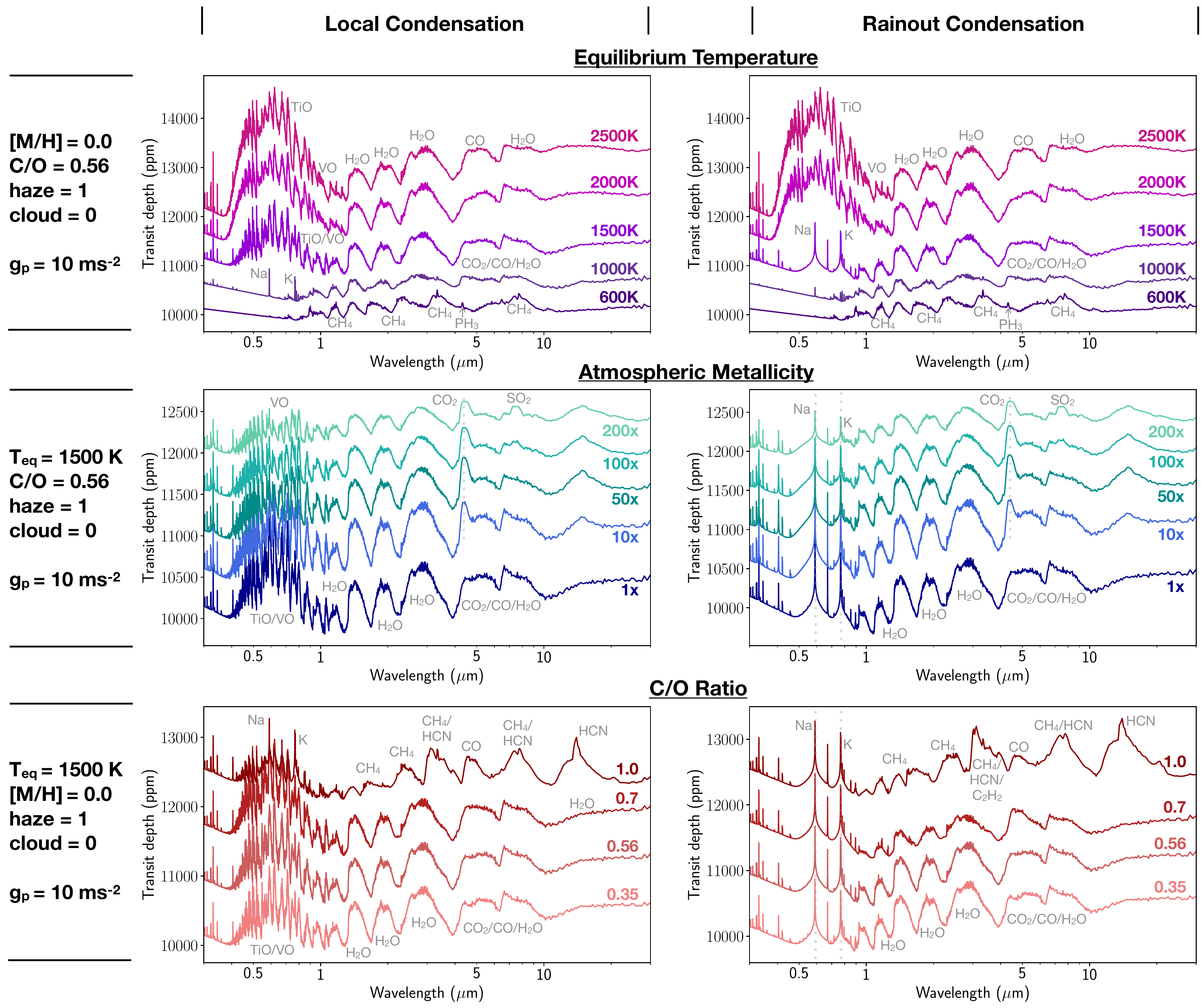}
    \caption{Figure showing transmission spectra with the effects of local condensation (left) and rainout condensation (right). \corr{We show the effects of varied temperature (top row) with a fixed metallicity and C/O ratio set to solar, varied metallicity (middle row) with a fixed T$_{eq}$\,=\,1500\,K and C/O at solar, varied C/O ratio (bottom row) with a fixed T$_{eq}$\,=\,1500\,K and solar metallicity.} In each case we consider a clear atmosphere with haze and cloud parameter set to 1.0 and 0.0, respectively, for a planet with radius 1 R$_{\rm J}$ around a star with radius 1 R$_{\rm sun}$ and g$_p$=10\,ms$^{-2}$. An offset in transit depth has been added to each spectrum for clarity.}
    \label{fig:ggg_rainout_test}
\end{figure*}

In this paper we present a publicly available\footnote{\url{https://drive.google.com/open?id=1ZFbkPdqg37_Om7ECSspSpEp5QrUMfA9J}} new generic grid of forward model simulations that can be scaled to a wide range of star-planet pairs for atmospheric transmission spectra. \corr{We provide a user-friendly generic grid of simulated transmission spectra to interpret observations, where the word \enquote{generic} implies a grid of models that can be scaled to a wide range of H$_2$/He dominated exoplanet atmospheres. We highlight the sensitivity of our model transmission spectra to choices in atmospheric physics, such as condensation schemes. We also provide comparisons between our scalable grid and planet specific grid as well as spectra generated by other atmospheric modeling frameworks. These comparisons critically highlight how choices made in generating atmospheric models influence our interpretation of the underlying physics captured by observations.} In Section \ref{sec:model} we first detail the model, its setup for the grid and treatment of condensation. \corr{In Section \ref{sec:gridparamspace} we describe the parameter space of the grid. In Section \ref{sec:effectofdifferentparameters} we present the scientific results obtained from the grid by detailing the effects of sensitivity tests on atmospheric chemical composition and the resultant transmission spectra.} In Section \ref{sec:usinggrid} we detail how the models can be scaled to any planet-star combination and compare them with other published model grids in the literature for validation. In Section \ref{sec:transindex} we discuss the application of the grid with specific reference to the transmission spectral index established in \citet{Sing2016} and finally we conclude in Section \ref{sec:conclusions}. 

\section{\texttt{ATMO} Forward Models}
\label{sec:model}
We use \texttt{ATMO}, a 1D-2D radiative-convective equilibrium model for planetary atmospheres \citep{Tremblin2015,Tremblin2016,Amundsen2014,Drummond2016,Tremblin2017,Goyal2018} to compute a grid of generic forward models, which can be scaled to represent a wide range of H$_2$/He dominated atmospheres. For this work we use isothermal $P$-$T$ profiles under the assumption of chemical equilibrium. We include H$_2$-H$_2$ and H$_2$-He collision induced absorption (CIA) opacities. We also include opacities due to H$_{2}$O, CO$_2$, CO, CH$_4$, NH$_3$, Na, K, Li, Rb, Cs, TiO, VO, FeH, CrH, PH$_3$, HCN, C$_{2}$H$_{2}$, H$_{2}$S and SO$_{2}$. The source of these opacities and their pressure broadening parameters can be found in \citet{Amundsen2014} and \citet{Goyal2018}. We note that in this work we adopt Na and K pressure broadened line profiles as derived in \citet{Burrows2000}, instead of \citet{Allard2003}, adopted for planet specific grid presented in \citet{Goyal2018}. This was motivated by the results of \citep{Nikolov2018} where the profiles of \citet{Burrows2000} had a slightly better fit to observations than that of 
\citet{Allard2003}, although the final results were statistically inconclusive while comparing these different pressure broadened profiles. 

This grid of model simulations is baselined for a Jupiter radius planet (1 R$_{\rm J}$ at 1 millibar pressure) around a Solar radius star (1 R$_{\rm sun}$). Our isothermal $P$-$T$ profiles extend from $10^{-6}$ bar at the top of the atmosphere to 10 bar at the bottom, with the radius of the simulated planet that is 1 R$_J$, defined at the 1 millibar pressure level, which approximates the region of the atmosphere probed with transmission spectra \citep{Lecavelier2008}. \corr{The upper and the lower boundary conditions of the atmosphere, for the equilbrium chemical abundances calculation are set by the input pressure grid. We use 50 model levels for each isothermal $P$-$T$ profile, which are evenly spaced in log(P) space. Over a large pressure range the assumption of an isothermal atmosphere is an extreme assumption, especially for isothermal $P$-$T$ profiles with rainout (see Section \ref{sec:rainout}). Therefore, we set the bottom of the atmosphere pressure to 10 bar, approximately similar to previous works \citep[e.g][]{Fortney2010}.} We compute transmission spectra in 5000 correlated-k bins, evenly spaced in wavenumber, which corresponds to R$\sim$5000 at 0.2 \textmu m while decreasing to R$\sim$100 at 10 \textmu m \citep[see][for details]{Goyal2018}.

\subsection{The Treatment of Condensation and Rainout}
\label{sec:rainout}

Within our model we adopt three different approaches to compute chemical equilibrium abundances; gas-phase only, local condensation and rainout condensation. We do not produce a grid for the gas phase only approach, as it can lead to unphysical spectral features for planetary atmospheres, where condensates can form. 
In the \emph{local condensation} (no-rainout) approach, each model level is entirely independent of all other model levels along the profile, and the chemical composition is only dependent on the local temperature and pressure, and the element abundances. In this approach, each model level starts with the same elemental abundance, and condensates that form, only deplete the material in that specific layer of the atmosphere.

In contrast, in the condensation with \emph{rainout} approach, once the condensates are formed, the elements that comprise that condensate are depleted stoichiometrically from the concerned layer and all the layers above it. This is driven by the assumption that once condensates are formed they will tend to sink in the atmosphere (a limiting case), and the elements that compose those condensate species will therefore be depleted for lower pressures. For example, formation of Mg$_2$SiO$_4$ condensate will remove 2:1:4 of the Mg, Si and O atoms in the layers above\corr{, this will continue until one or more of the elements are depleted in their entirety.} This setup is designed to mimic cloud particles forming and \enquote{settling} or \enquote{sinking} in an atmosphere (i.e. raining out) \citep{Burrows1999,Mbarek2016}. Therefore, in the rainout approach each layer is dependant on the deeper (high pressure) layers of the atmosphere for its chemical abundances. We note that, since O is one of the most abundant element that gets locked up in the condensates, rainout processes
primarily effect the availability of O in the atmosphere for the upper layers. For a detailed description of the implementation of the rainout calculation in our model, see Section 3.3.2 of \citet{drummond2017}\footnote{\url{https://ore.exeter.ac.uk/repository/handle/10871/27993}}. 

$P$-$T$ profiles in radiative-convective equilibrium are substantially hotter in deeper levels (>10 or 100 bar) compared to upper region (<10 bar) of the atmosphere (except when there is an inversion). Since the rainout process is dependant on the adopted $P$-$T$ profile, assumption of isothermal $P$-$T$ profile in deeper levels of the atmosphere can be an extreme assumption, as it would form more condensates compared to a hotter radiative-convective equilibrium $P$-$T$ profile and thereby effect the chemical abundances in the upper region (low pressure region) of the atmosphere probed by transmission spectra. Therefore, we adopt 10 bars as the lower boundary of our model domain in this work, which is generally the deepest part of the atmosphere, transmission spectra can probe \citep{Fortney2010}.
These two different condensation approaches can have large effects on the chemical equilibrium abundances of various molecules across different parameter spaces and therefore the transmission spectra. Therefore, to make the computed grids applicable to differing scenarios we produce a generic grid for both local condensation and rainout condensation conditions. We note that the assumption of isothermal $P$-$T$ profiles is not as accurate as $P$-$T$ profiles in radiative-convective equilibrium \citep[e.g.][Goyal et al. in prep.]{Molliere2016}. However, isothermal models have been shown to be sufficient to interpret current transmission spectra observations \citep{Fortney2005, Heng2017, Goyal2018}, and are much more computationally efficient. 

\begin{figure*}
	\includegraphics[width=\textwidth]{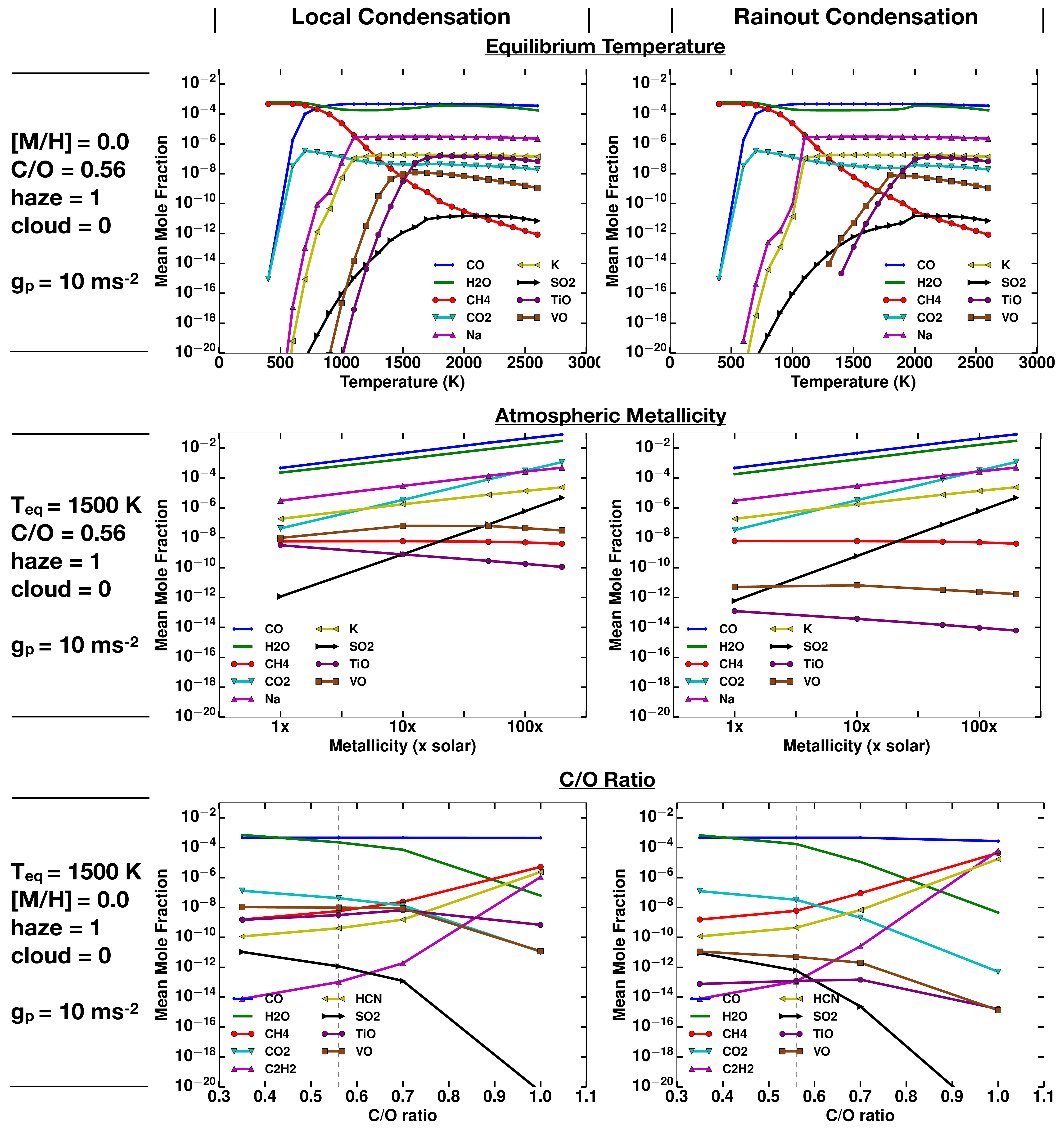}
    \caption{Figure showing change in mean chemical abundances between 0.1 and 100 millibar for various molecules for the same model simulations as in Fig. \ref{fig:ggg_rainout_test}. Changes in mean chemical abundances due to  local condensation (left) and rainout condensation (right)  \corr{for varied temperature (top row) with a fixed solar metallicity and C/O ratio, varied metallicity (middle row) with a fixed T$_{\rm eq}$\,=\,1500\,K and solar C/O ratio, and for varied C/O ratios with a fixed T$_{\rm eq}$\,=\,1500\,K and solar metallicity (bottom row)].}}
    \label{fig:ggg_rainout_chem}
\end{figure*}
\section{Grid Parameter Space}
\label{sec:gridparamspace}
Our aim is to produce a forward model grid which can be scaled to any planet-star pair for both, local condensation and rainout condensation cases. The grid requires a sizeable number of parameters which must be sampled with sufficient fineness to allow accurate interpolation. For each of the two treatments of condensation we compute \corr{28,160} forward models over a range of 22 planetary equilibrium temperatures (400, 600, 700, 800, 900, 1000, 1100, 1200, 1300, 1400, 1500, 1600, 1700, 1800, 1900, 2000, 2100, 2200, 2300, 2400, 2500, 2600; all listed in K), four planetary surface gravities (5, 10, 20, 50; all listed in ms$^{-2}$), five atmospheric metallicities (1, 10, 50, 100, 200; all in $\times$\,solar), \corr{four C/O ratios (0.35, 0.56, 0.7, 1.0)}, four scattering haze parameters (1, 10, 100, 1100$\times$ standard Rayleigh-scattering) and four uniform cloud parameters (0, 0.06, 0.2, 1). Haze is treated as a small scattering aerosol particles and implemented as a parameterized enhanced multi-gas Rayleigh scattering, while clouds are treated as large particles with grey opacity. \corr{Haze parameterization is implemented via the equation $\sigma(\lambda)=\alpha_{haze}\sigma_{0}(\lambda)$ where $\sigma(\lambda)$ is the total scattering crossection with haze, $\alpha_{haze}$ is the haze enhancement factor and $\sigma_{0}(\lambda)$ is the scattering crossection due to all other gases. The cloud parameterization is implemented vi the equation $\kappa(\lambda)_c = \kappa(\lambda) + \alpha_\textup{cloud}\kappa_{H_2}$, where $\kappa(\lambda)_c$ is the total scattering opacity in $\textup{cm}^{2}/\textup{g}$, $\kappa(\lambda)$ is the scattering opacity due to nominal Rayleigh scattering in same units, $\alpha_\textup{cloud}$ is the variable cloudiness factor governing the strength of grey scattering and $\kappa_{H_2}$ is the scattering opacity due to H$_2$ at 350 nm which is $ \sim 2.5 \times 10^{-3} \textup{cm}^{2}/\textup{g}$ \citep[see][for more details]{Goyal2018}.} 

\corr{We do not extend the grid to sub-solar metallicities and limit the high metallicity end to 200$\times$\,solar. Above 200$\times$\,solar the atmosphere becomes abundant in species other that H$_2$ and He, such as CO$_2$, H$_2$O, CO etc. This would require the inclusion of pressure broadening effects due to these species, and no existing studies have solved this problem due to lack of lab-based observational data \citep{Fortney2016}. The choice of C/O ratios in the grid is guided by important transition regimes as found by previous studies \citep{Madhusudhan2012, Molliere2015, Goyal2018}. Each of the cloud and haze parameters apply a scaling from 0.0 or 1$\times$, respectively, up to extreme values of almost total obscuration (1.0) for uniform clouds and 1100$\times$ the scattering parameter for haze.}

\section{\corr{Sensitivity Analysis of the Grid}}
\label{sec:effectofdifferentparameters}
\corr{This model grid explores a wide range of giant planet parameter space across temperatures and chemical abundances, for two different condensation regimes. Therefore, we detail the sensitivity of spectra to different parameter and physics choices in this section.
We show a subset of our simulations for both local condensation (left) and rainout condensation (right) in Fig. \ref{fig:ggg_rainout_test}, demonstrating the changes with temperature (top), metallicity (middle) and C/O ratio (bottom). For the same 3 grid parameters, we also show the changes in the mean chemical abundances of various spectrally important species in Fig. \ref{fig:ggg_rainout_chem}.} 

\subsection{\corr{Impact of changing temperature}} 
Changes in temperature have a more profound impact on the transmission spectrum compared to any other parameter, as the chemical composition is strongly dependent on the temperature, \corr{under the assumption of chemical equilibrium \citep{Burrows1999}. To demonstrate the impact of temperature on the transmission spectrum we fix the metallicity and C/O ratio to solar value and use a clear atmospheric (haze and cloud parameters of 1.0 and 0.0, respectively) spectrum for a planetary gravity of 10\,ms$^{-2}$, as shown in Fig. \ref{fig:ggg_rainout_test}(top row)}. In both condensation cases, at low temperatures <1000\,K the spectra are dominated by CH$_4$ with minor contributions from H$_2$O. Above $\sim$1000\,K the spectra are dominated by H$_2$O absorption with contributions in the IR from CO and CO$_2$. This can also be noticed in Fig. \ref{fig:ggg_rainout_chem} bottom row, where the CH$_4$ abundance drops dramatically after $\sim$1000\,K. The main difference between the two condensation cases can again be seen in the optical. In the local condensation case TiO/VO absorption is expected at temperatures above $\sim$1400\,K, while in rainout condensation, absorption by TiO/VO is suppressed until $\sim$1700\,K. This \corr{implies} that the atomic Na and K line absorption is only apparent in the local condensation grid between $\sim$ 800--1400\,K. In the rainout condensation grid, the presence of Na and K is also impacted by condensation, therefore Na and K features are only expected between $\sim$ 1100--1700\,K

\subsection{\corr{Impact of changing metallicity}} 
\corr{The effects of metallicity are shown for a clear atmosphere at an equilibrium temperature of 1500\,K, solar C/O ratio(0.56) and gravity of 10\,ms$^{-2}$}. Generally, it can be seen that all the spectral features tend to be reduced in amplitude with increase in metallicity as seen in Fig. \ref{fig:ggg_rainout_test} (middle row), although the abundance of all the spectrally important species increase, as seen in Fig. \ref{fig:ggg_rainout_chem}. This is caused by the increasing metallicity leading to a reduction in the scale height of the atmosphere, as the mean molecular weight increases. With increase in metallicity, there is also increase in CO$_2$ abundance \citep{Moses2013b} as seen in Fig. \ref{fig:ggg_rainout_chem}, which can also be noticed by rise in strong CO$_2$ feature at 15\,\textmu m. 

In the local condensation case, the strength of TiO/VO features in the optical decreases with increasing  metallicity at 1500\,K, which is a combination of decrease in TiO/VO abundance as well as muting of the features due to increased molecular weight of the atmosphere. In contrast, rainout condensation entirely removes TiO and VO features at 1500\,K due to condensation and removal of the species deeper in the atmosphere, leading to their lower abundances. The lower abundances of TiO and VO, leaves the atomic lines of Na and K as the dominant transmission features in the optical. There is, however, little difference (between rainout and no-rainout) in the infrared (IR) with only moderate changes in the absolute abundance of the oxygen based species due to condensate formation. The IR spectra are mainly dominated by H$_2$O features. However, CO and CO$_2$ features can be seen between 4 and 5\,\textmu m. 
Additionally, there is a SO$_2$ feature between 7 to 8\,\textmu m which first appears in the 100$\times$ solar metallicity models, with a stronger feature at 200$\times$ solar metallicity. The presence of SO$_2$ can possibly be used to constrain the metallicity of exoplanet atmospheres with mid-IR observations and strong near-IR constraints on the H$_2$O abundance. \corr{SO$_2$ is one of the most prominent sulphur gases in the atmosphere of Venus, having substantial effect on its radiative balance \citep{Vandaele2017}.  Therefore, investigating the possibility of SO$_2$ detection, can increase our understanding of the sulphur cyle in giant planetary atmospheres.}

\subsection{\corr{The impact of changing C/O}}
\corr{The effects of varying the carbon-to-oxygen (C/O) ratio are shown in Fig. \ref{fig:ggg_rainout_test} and \ref{fig:ggg_rainout_chem} (bottom row) for a clear atmosphere at an equilibrium temperature of 1500\,K, solar metallicity and gravity of 10\,ms$^{-2}$. The C/O ratio primarily dictates the dominance of various carbon and oxygen bearing molecular species in the atmosphere and thereby the spectra. As seen in Fig. \ref{fig:ggg_rainout_chem} bottom row, for C/O ratios less than equal to solar values, atmosphere is dominated by CO, H$_2$O and CO$_2$. However, for C/O ratios greater than solar, the abundances of molecules with carbon but without oxygen such as CH$_4$, HCN, and C$_2$H$_2$  start increasing rapidly. This effect is visible in the transmission spectra as shown in Fig. \ref{fig:ggg_rainout_test} (bottom row), where the infrared spectra is primarily dominated by H$_2$O features upto solar C/O ratio (0.56), 0.7 being the transitional value and 1 being the C/O ratio at which the infrared spectra is primarily dominated by CH$_4$ and HCN features. Choice of condensation also has an effect on spectral features, especially for C/O ratios greater than solar. It can be seen in Fig. \ref{fig:ggg_rainout_test} that at C/O ratio of 1, spectral features are different between rainout and local condensation cases. This is because the rainout case has a slighly larger abundance of CH$_4$, HCN and C$_2$H$_2$, as compared to local condensation case seen in Fig. \ref{fig:ggg_rainout_chem}. It can also be seen from this figure that the abundances of spectrally important molecules such as H$_2$O, C$_2$H$_2$ etc. are more strongly dependent on the C/O ratio in the rainout case, compared to local condensation case.}


\subsection{\corr{The presence of VO without TiO}} 
Recent observations have shown the possibility of VO with the absence of TiO in the atmosphere of extremely irradiated hot Jupiter WASP-121b (Equilibrium temperature (T$_{\rm eq}$)\,=\,2358\,K) \citep{Evans2017}. Our models suggest that this may only be possible across a narrow range of temperatures, namely $\sim$ 1200--1400\,K, under the assumption of local condensation as seen in Fig. \ref{fig:ggg_rainout_chem} (top row). In this rather narrow temperature regime, the abundance of VO is higher than that of TiO and sufficient enough to impart its features in the transmission spectrum. This is because the primary Ti condensate Ti$_3$O$_5$ is more abundant than the V condensate V$_2$O$_3$, thereby locking more Ti in condensates than V at these temperatures. For temperatures higher than 1400\,K TiO starts dominating the optical spectrum. However, for the rainout condensation case, this narrow range where VO is present without TiO, is $\sim$ 1700--1800\,K. Therefore, presence of VO features in the spectrum without TiO can help constrain the limb temperature of the planet's atmosphere, if the rainout and local condensation processes in the planetary limb are constrained. Additionally, if the planetary limb temperature is constrained using the Rayleigh scattering slope, TiO or VO features can reveal which process is dominant in these atmospheres (rainout or local condensation).


\section{Working with the Grid}
\label{sec:usinggrid}
\begin{figure}
	\includegraphics[width=\columnwidth]{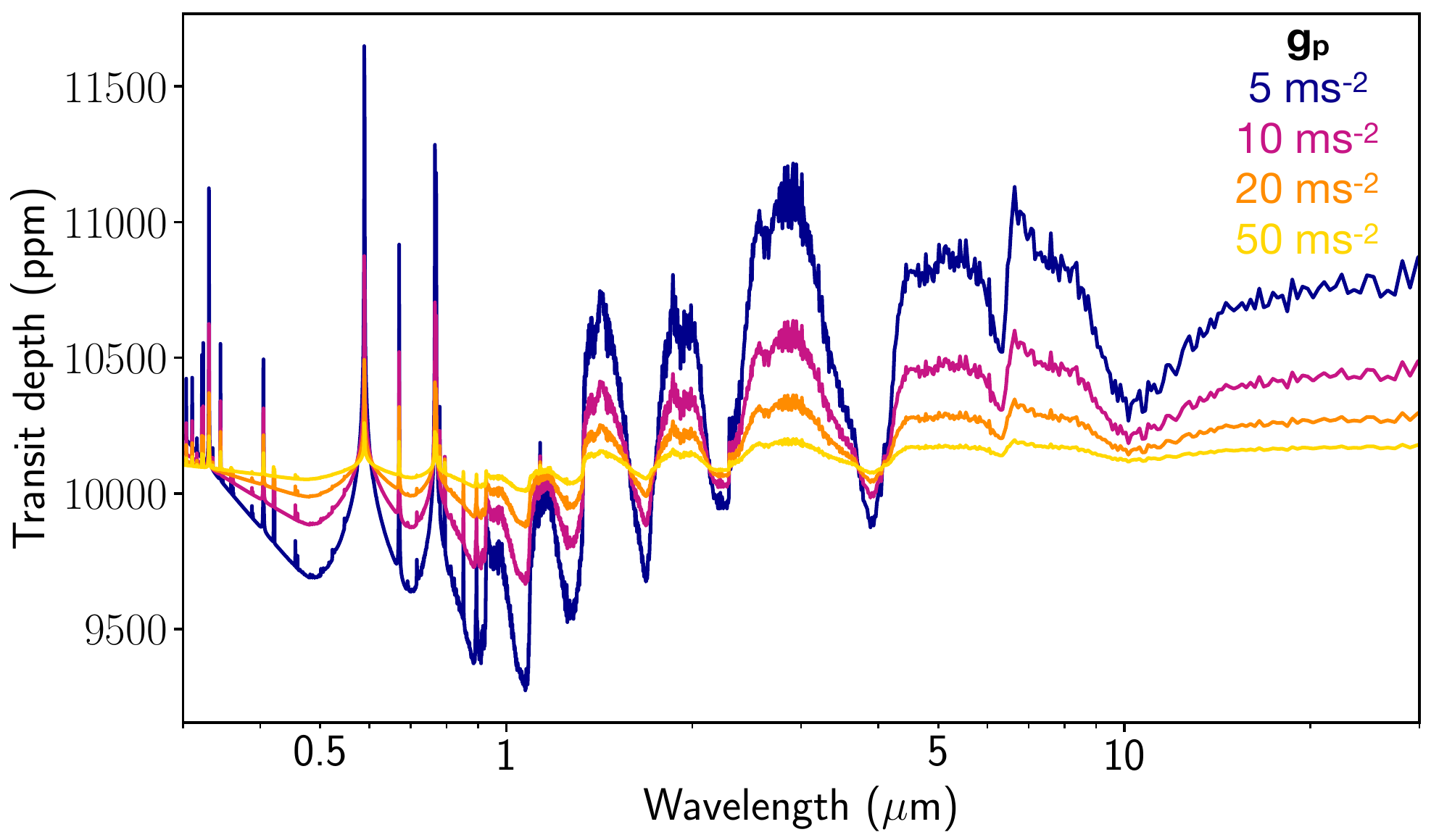}
    \caption{\corr{Figure showing how changing the gravity changes the size of the absorption features in the transmission spectrum, due to a change in the scale height, for each of the gravities represented in the grid (5\,ms$^{-2}$, 10\,ms$^{-2}$, 20\,ms$^{-2}$, 50\,ms$^{-2}$ ). Each model is for a clear solar metallicity ans solar C/O ratio atmosphere with T$_{eq}$\,=\,1500\,K.}}
    \label{fig:ggg_diff_gravity}
\end{figure}

The generic exoplanet \texttt{ATMO} model grid has been produced such that it can be scaled to a wide range of planet/star combinations and can be applied as an interpretive tool as well as a preparation tool for exoplanet atmospheric studies. 
Temperature and gravity are two of the most important parameters shaping the transmission spectrum of giant exoplanets as they effectively control the scale height of the atmosphere. The temperature also governs the chemical state of the atmosphere, where different temperatures can lead to different chemical properties (see Fig. \ref{fig:ggg_rainout_chem}). As such, the parameter space of the temperature (400-2600\,K) is broken down into fine bins (of $\sim$100\,K). The gravity, however, can be represented over fewer values and scaled to a more precise value. \corr{The amplitude of features in the transmission spectra are strongly tied to the scale height of the planet's atmosphere, which is inversely proportional to the planet's gravity as shown in Fig. \ref{fig:ggg_diff_gravity}. In short, all else remaining equal, as the gravity increases, the amplitude of spectral features decrease.} 

\begin{figure}
	\includegraphics[width=\columnwidth]{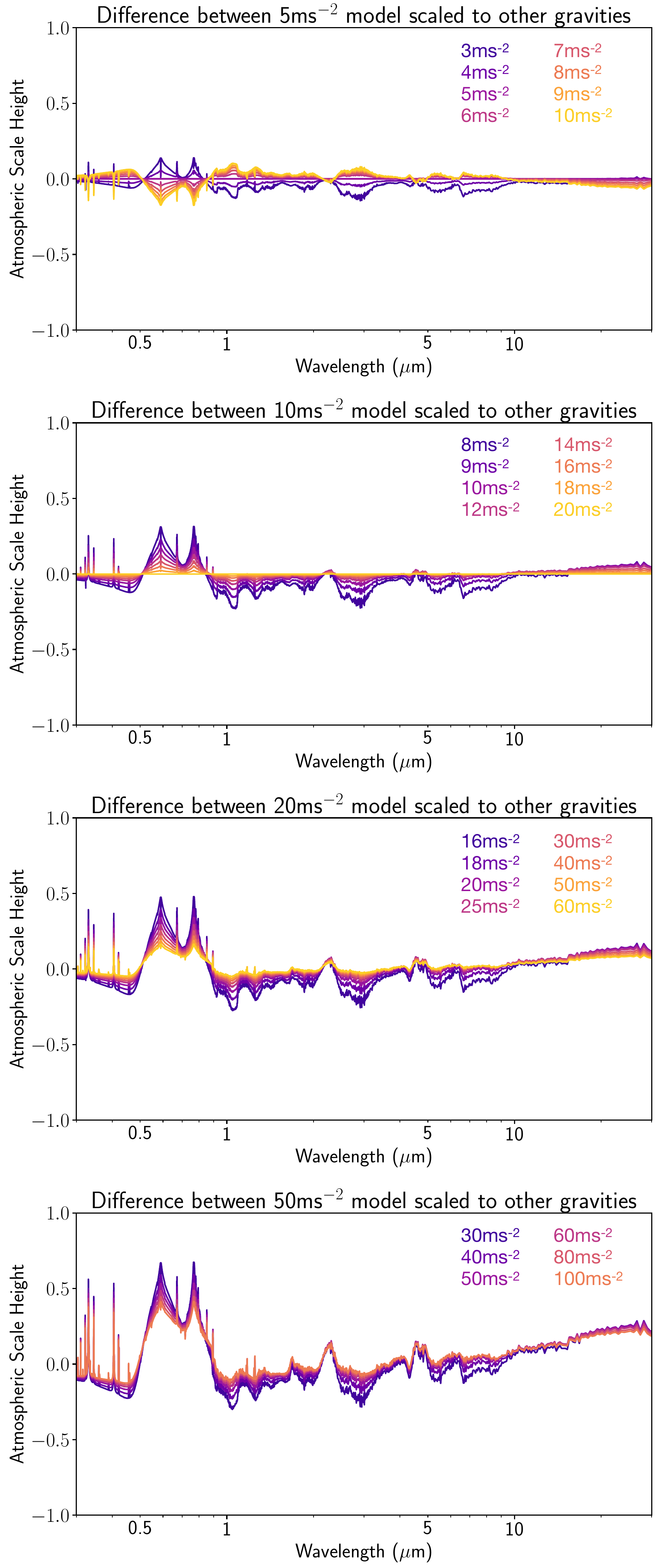}
    \caption{Figure showing the residual comparison of the generic model scaled to different planetary gravities to models computed at the specific gravities, for each of the four planetary gravity parameters in the grid (5\,ms$^{-2}$ top, 10\,ms$^{-2}$ middle top, 20\,ms$^{-2}$ middle bottom, 50\,ms$^{-2}$ bottom).}
    \label{fig:ggg_gravity_test}
\end{figure}

To demonstrate the scalability of the gravity, for each gravity parameter in the grid (5, 10, 20, 50\,ms$^{-2}$) we scale the model to a variety of different planetary gravities and compare it to a model specifically generated for that gravity. In Fig. \ref{fig:ggg_gravity_test}, we demonstrate the accuracy associated with scaling the four gravities supplied in the grid to a finer parameter space between 3--100\,ms$^{-2}$. \corr{The residuals are well below half the scale height of the atmosphere, when scaled from the 5, 10 or 20\,ms$^{-2}$ models as seen in Fig. \ref{fig:ggg_gravity_test}. However, residuals tend to increase beyond half scale height for 50 \,ms$^{-2}$ models. One of the reasons for this is the difference between the gravity values ($\delta$g) used for scaling, for lower values of g smaller $\delta$g is used, thus resulting in smaller residuals and for higher values, larger $\delta$g results in comparatively larger residuals. It can also be noticed that the residuals are maximum between 0.5 and 1\,\textmu m, specifically around Na and K absorption bands at 0.58 and 0.76\, \textmu m respectively. This is because when transmission spectra is computed for a specific value of gravity it corresponds to specific value of pressure and therefore specific pressure broadening of opacities. However, scaling to a different gravity changes the pressure level probed by transmission spectra, since it varies as square root of g (see Equation \ref{eq:lecav}) and therefore ideally requires pressure broadened opacities at this new pressure levels. But while scaling generic models between different gravities, we don't alter the opacities, leading to largest residuals for Na and K bands which are very strongly affected by pressure broadening.}  


The other parameters covered by the grid represent different scaling parameters applied to either the chemistry (e.g. metallicity, \corr{C/O ratio}) via a scaling of the abundances, or in the opacity (e.g. haze and cloud, via scaling factors with and without wavelength dependence; see \citealt{Goyal2018} for details). 
 
\subsection{Scaling to Specific Planetary Parameters}
\label{sec:scaling}
In order for these models to be applied to the desired exoplanet they will have to be scaled based on the planetary radius, stellar radius, surface gravity, and planetary equilibrium temperature. 

The wavelength-dependent observed (apparent) radius of the planet $R_{p}({\lambda})$ can be seen as a combination of the wavelength-independent bulk planet radius $R_{p,bulk}$ and a wavelength-dependent contribution from the atmosphere $z(\lambda)$,

\begin{equation}
R_{p}(\lambda) = z(\lambda) + R_{p,bulk} \textrm{,}
\label{eq:rp} 
\end{equation}

where the term $z(\lambda)$ is dependent on the physical and chemical properties of the atmosphere. 

\citet{Lecavelier2008} provide an approximate analytic solution for $z(\lambda)$, 

\begin{equation}
z(\lambda) = H\ln\Bigg(\frac{\xi_{abs}P_{z=0}\sigma_{abs}(\lambda)}{\tau_{eq}}  \sqrt{\frac{2 \pi R_{p,bulk}}{k_{b} T \mu g}} \Bigg) = H\ln\alpha \textrm{,}
\label{eq:lecav} 
\end{equation}

where $T$ is temperature, $\mu$ is the mean molecular weight, $g$ is gravity, $k_b$ is the Boltzmann constant and $H = (k_{b} T)/(\mu g)$ is the atmospheric scale height. $\sigma_{abs}(\lambda)$ and $\xi_{abs}$ are the absorption cross section and mole fraction of the dominant absorbing species, respectively. $P_{z=0}$ is the pressure at the effective altitude($z$) of 0 (i.e at the base of the atmosphere). Optical depth, $\tau_{eq}$, is set to 0.56 \citep{Lecavelier2008}. We note that we do not use Eq. \ref{eq:lecav} to calculate the transmission spectra in our model grid, but use the numerical approach detailed in \citet{Amundsenthesis,Goyal2018}.

In our model grid we present a large number of transmission spectra for specific sets of model parameters, where, $R_{p}^{grid}(\lambda) = z^{grid}(\lambda) \, + \,  R_{p,bulk}^{grid}$, as shown in Eq. \ref{eq:rp}.  However, we
derive a scaling relation by solving Eq. \ref{eq:lecav} simultaneously for grid and planet parameters, to fine-tune a particular model from the grid to a specific set of planetary parameters, 

\begin{equation}
\frac{z^{grid}(\lambda)}{H^{grid}} - \frac{z^{planet}(\lambda)}{H^{planet}} = \ln\alpha^{grid} - \ln\alpha^{planet}
\label{eq:simul}
\end{equation}

where terms denoted \enquote{grid} are values from the model grid and terms denoted \enquote{planet} are the new parameters for a specific case. Rearranging Eq. \ref{eq:simul} and canceling constants we obtain the scaling relation,

\begin{equation}
z^{planet}(\lambda) =  z^{grid}(\lambda)\frac{T^{planet}g^{grid}}{T^{grid}g^{planet}} - 0.5\ln\frac{R_{p,bulk}^{grid}T^{planet}g^{planet}}{R_{p,bulk}^{planet}T^{grid}g^{grid}} \textrm{.}
\label{eq:final}
\end{equation}

Importantly, we note that we have made the assumption that $\sigma_{abs}$, $\xi_{abs}$ and $\mu$ are constants, while scaling from the nearest grid to planetary parameter, which is a reasonable assumption given the fine variation of parameters in the grid, as demonstrated in Section \ref{sec:planetspecific}

The wavelength dependent planetary radius ($R_{p}({\lambda})$)  scaled to the parameters of a specific planet can then be found using Eq. \ref{eq:rp} and Eq. \ref{eq:final}. The transmission spectrum $ \Big (\frac{R_{p}({\lambda})}{R_*} \Big)$ can then be obtained simply by including the relevant stellar radius $R_*$. We provide a python code on GitHub and the grid google drive to scale these models\footnote{\url{https://github.com/hrwakeford/Generic\_Grid}}$^,$\footnote{\url{https://drive.google.com/open?id=1ZFbkPdqg37_Om7ECSspSpEp5QrUMfA9J}}.

\subsection{Comparison to Planetary Specific Grid}
\label{sec:planetspecific}
\begin{figure}
	\includegraphics[width=\columnwidth]{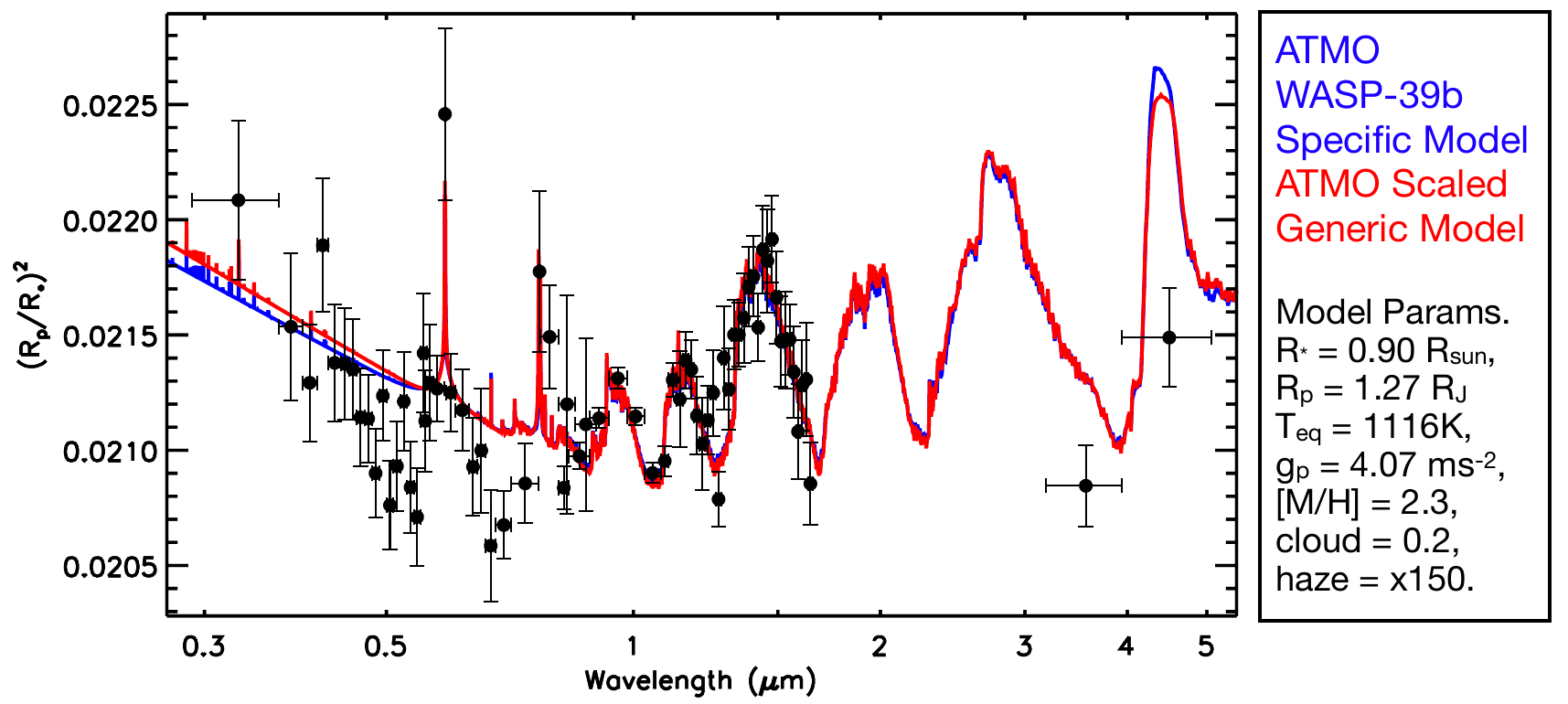}
    \caption{Figure showing the use of the generic grid (red) to fit atmospheric data (black points, \citealt{Wakeford2018}), in comparison to the planetary forward model grid (blue) presented in \citet{Goyal2018}. In both cases we use the grid for local condensation only.}

    \label{fig:W39_test}
\end{figure}
We test this new scalable \texttt{ATMO} grid by comparing it to the previously published planetary specific transmission spectra \citep{Goyal2018}\footnote{\url{https://bd-server.astro.ex.ac.uk/exoplanets/}}. WASP-39b has T$_{\rm eq}$\,=\,1116\,K, g$_{\rm p}$\,=\,4.07\,ms$^{-2}$, R$_{\rm p}$\,=\,1.27\,R$_{\rm J}$, and R$_{*}$\,=\,0.90\,R$_{\rm sun}$ adopted from TEPCAT database \citep{Southworth:2011aa}. As shown in \citet{Wakeford2018}, the best fit model for WASP-39b data from the \citet{Goyal2018} grid has a metallicity of 100$\times$\,solar, uniform cloud = 0.2, and scattering haze = 150$\times$. To demonstrate the applicability of our generic model spectra grid for different planetary systems, we scale to the planetary parameters of WASP-39b, and compare to this best fit model in Fig. \ref{fig:W39_test}. From the generic grid we scale the T$_{\rm eq}$\,=\,1100\,K,  g$_{\rm p}$\,=\,5.0\,ms$^{-2}$, 200$\times$\,solar, uniform cloud = 0.2, and scattering haze = 150$\times$ model to the best fit model parameters for WASP-39b. For this temperature range the spectrum is relatively independent of the treatment of the condensation, so for clarity we only show the local condensation spectrum in Fig.\,\ref{fig:W39_test}.

This test demonstrates the flexibility of this grid of models to be adapted to specific planetary parameters. There are minor differences, $\sim$100ppm, in the IR near 4.5 microns with the scaled CO/CO$_2$ feature, however, these do not significantly affect the fit to the data. In the rest of the spectra, the differences between the models average 50 ppm, which is well below the data uncertanties. The minor differences in the Rayleigh scattering slope can be attributed to differences in adopted pressure broadening profiles for Na in this work and that in \citep{Goyal2018}, explained in detail in Section \ref{sec:model}.

This test also demonstrates that differences shown in Fig. \ref{fig:ggg_gravity_test} when scaling the \corr{5\,ms$^{-2}$} model to new planetary parameters are negligible between specifically generated models and our re-scaled generic models, where differences are well within the constraints of the current observational measurements. 

\corr{Compared to the planetary specific grid presented in \citep{Goyal2018} this grid can be applied to  a wide range of H$_2$/He dominated exoplanet atmospheres and is not limited to the 117 well studied planets. This grid is applicable to targets detected using TESS, NGTS, HATS and any number of other H$_2$/He dominated planets discovered in the future over a wide parameter space. Due to the scalable nature of the grid it can also be implemented within a retrieval framework, for which the planetary specific grid is not suitable.}

\subsection{Comparison to other Forward Models}
\label{sec:othermodelcomp}
\begin{figure}
	\includegraphics[width=\columnwidth]{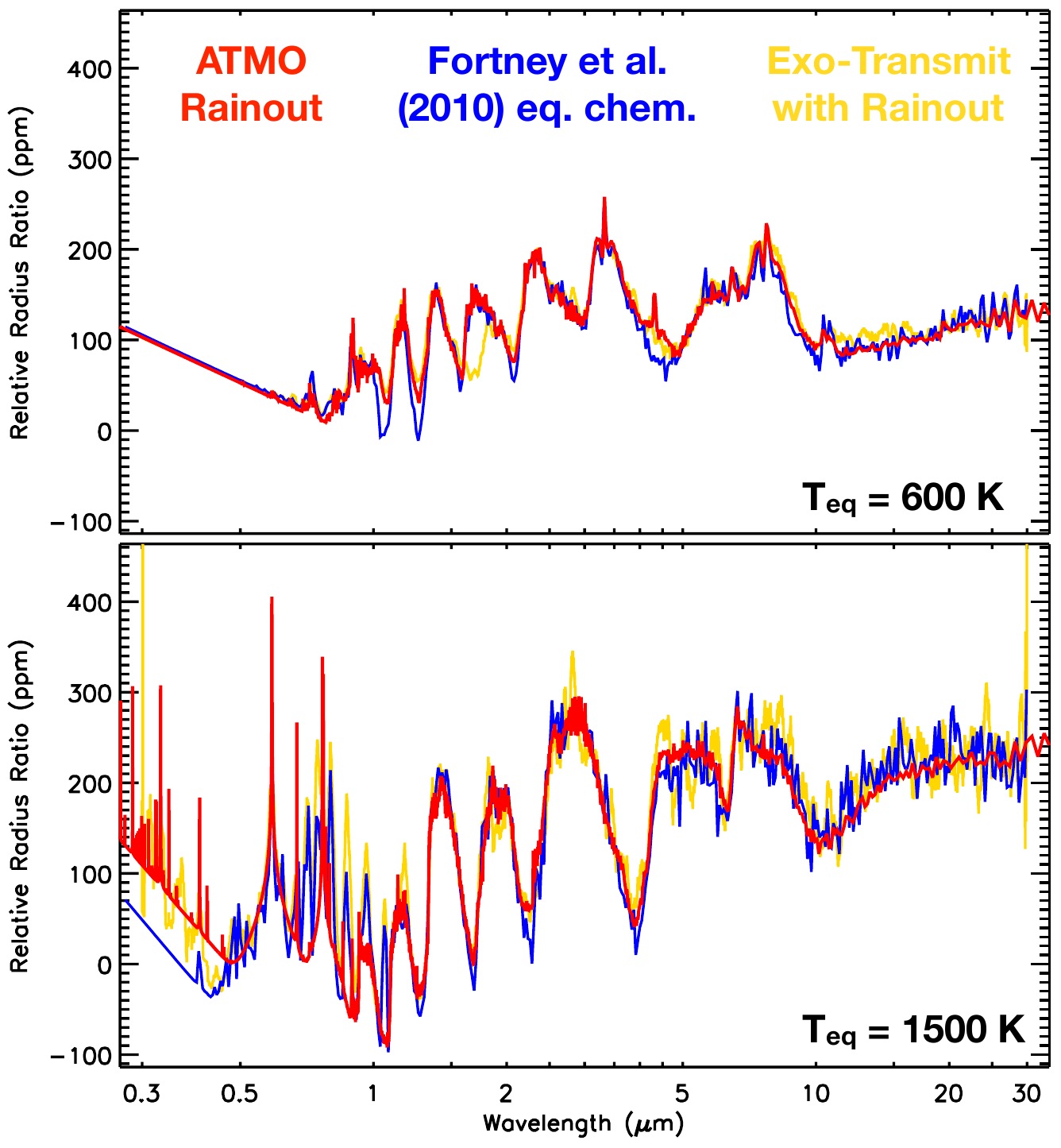}
    \caption{Figure showing comparison of \texttt{ATMO} rainout condensation model (red) to the \citet{Fortney2010} equilibrium model with TiO/VO (blue), and Exo-transmit with condensation and rainout (yellow). Each are isothermal models for g$_{\rm p}$=25\,ms$^{-2}$, R$_{\rm p}$/R$_*$=0.1, and T$_{\rm eq}$= 600\,K (top) and 1500\,K (bottom). We show the spectra aligned in the infrared 1.4\,$\mu$m. The main differences in the infrared are due to the opacity database used for H$_2$O and CH$_4$. Bottom: In the optical, Exo-transmit and the \citet{Fortney2010} show evidence for VO absorption while this is removed in the rainout process in \texttt{ATMO}.}
    \label{fig:ggg_vs_fg}
\end{figure}

Figure \ref{fig:ggg_vs_fg} shows the \enquote{scaled} simulated spectrum from the generic exoplanet \texttt{ATMO} forward model grid, with similar generic forward model simulations from \citet{Fortney2008,Fortney2010} and Exo-transmit from \citet{Kempton2017}. Each simulation is computed for isothermal P-T profiles assuming thermochemical equilibrium. The simulations in Fig. \ref{fig:ggg_vs_fg} are computed for solar metallicity atmospheres without cloud opacities and T$_{\rm eq}$\,=\,600\,K (top), T$_{\rm eq}$\,=\,1500\,K (bottom), with g$_{\rm p}$\,=\,10\,ms$^{-2}$, R$_{\rm p}$\,=\,1\,R$_{\rm J}$, and R$_*$\,=\,1\,R$_{\rm sun}$. \corr{Both Exo-transmit \citep{Kempton2017} and the models based on \citet{Fortney2008,Fortney2010}, included thermochemical equilibrium scheme which account for condensation \citep{Lodders1999,Lodders2002,Lodders2006,LoddersFegley2002,LoddersFegley2006,Visscher2006,Freedman2008}.} 

We see that all three models agree well at low temperature (T$_{\rm eq}$\,=\,600\,K) (with a slight difference in the ExoTransmit model between 1--2\,\textmu m). The difference in the relative radius ratio in the optical slopes at T$_{\rm eq}$\,=\,1500\,K between models from \citet{Fortney2010} and \texttt{ATMO}, can be attributed to raining out of TiO/VO in \texttt{ATMO} at these temperatures. 

Here we discuss some of the reasons for differences between these models as seen in Fig. \ref{fig:ggg_vs_fg}. Both Exo-transmit \citep{Kempton2017} and the models based on \citet{Fortney2008,Fortney2010} use elemental abundances from \citet{Lodders2003}. However, \texttt{ATMO} uses elemental abundances from \citet{Asplund2009}. The major difference between these two sources is in Helium abundance which is greater in \citet{Lodders2003} compared to \citet{Asplund2009}. This can lead to some differences in equilibrium chemical abundances. \corr{The differences in adopted polynomial coefficients for chemical equilibrium calculations can also lead to differences}. In \citet{Fortney2010} the base radius is either at 10 or 100 bar, while in this work we adopt 10 bar. The grid generated using ATMO uses high temperature line-lists primarily from Exomol with H$_2$/He pressure broadening applied wherever possible. In comparison, \citet{Fortney2010} and \citet{Kempton2017} models primarily use HITRAN line-lists. This can lead to differences in spectral features.

\section{Transmission Spectral Index}
\label{sec:transindex}
\begin{figure*}
	\includegraphics[width=\textwidth]{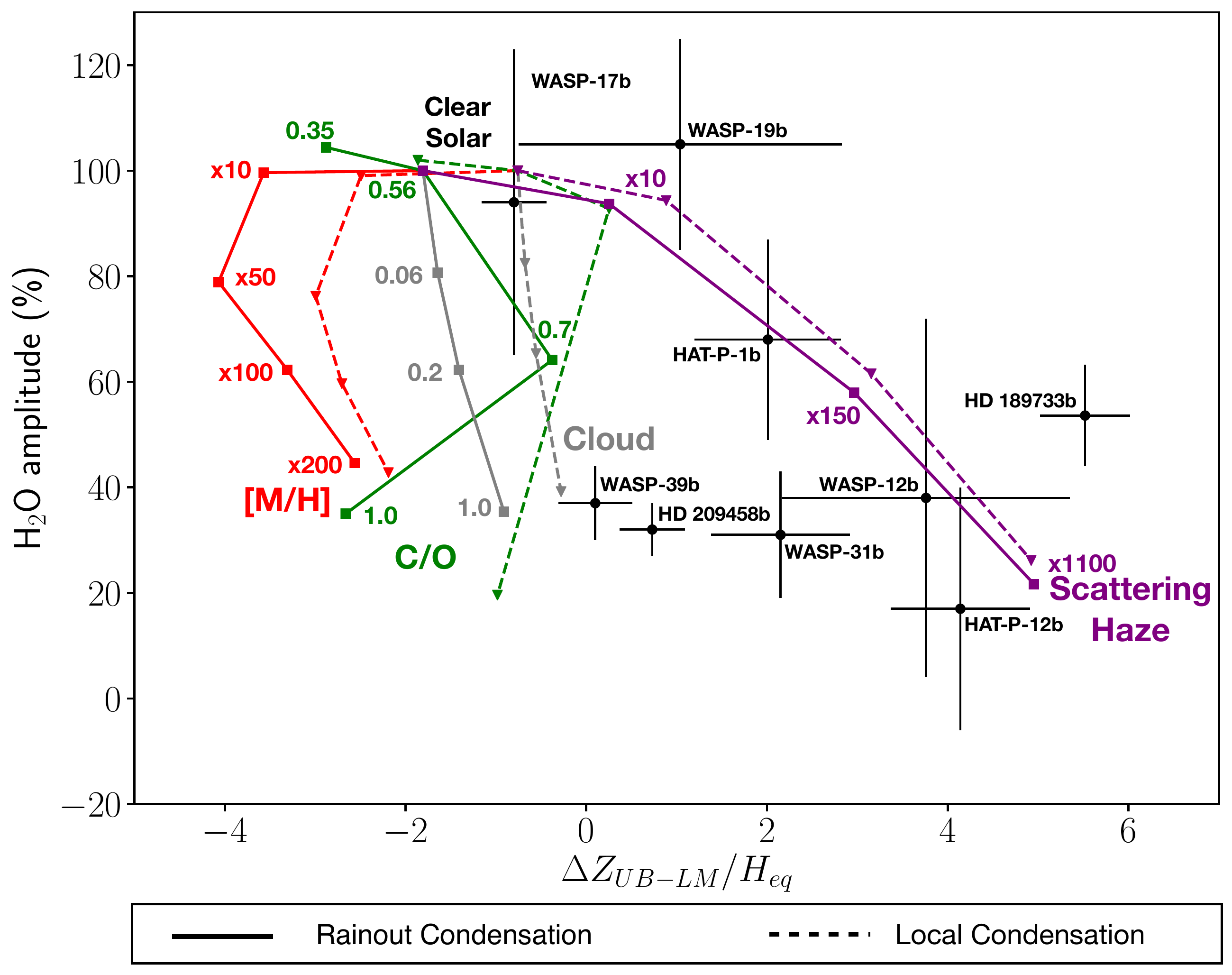}
    \caption{Figure showing \texttt{ATMO} grid  based transmission spectral index tracks first (e.g. \citet{Sing2016}). \corr{We show the H$_2$O amplitude versus the $\Delta z_{UB-LM}$ index for [M/H] (red), C/O (green), uniform cloud (grey), and scattering haze (purple) parameters}. We show the tracks for both the grid with rainout condensation (solid lines) and local condensation (dashed lines). Planetary data points are taken from \citet{Sing2016} and \citet{Wakeford2018}.}
    \label{fig:spectral_index}
\end{figure*}

To evaluate the impact that changing individual scaling parameters has on the atmospheric transmission spectrum we use the grid of generic models to compute the different transmission spectral indices as detailed in \citep{Sing2016}. We compute spectral indices for both local condensation and rainout condensation grids, and measure the difference in the radius ratio with an increase in atmospheric metallicity, uniform cloud scattering, and wavelength dependent haze scattering. We compute the H$_2$O amplitude versus $\Delta Z_{UB-LM}/H_{eq}$ index for both the grids and show the model trends in Fig. \ref{fig:spectral_index}. 

For each model, the H$_2$O amplitude is computed based on the radius ratio of each model between 1.34--1.49\,$\mu$m relative to the standard solar model over the same wavelength range. The relative change in the radius ratio for each model is then converted to a percentage amplitude of the H$_2$O feature compared to solar. 
The $\Delta z_{UB-LM}/H_{eq}$ index represents the measured radius ratio of the model in the blue-optical UB-band (0.3--0.57\,\textmu m) compared to the mid-infrared LM-band (3--5\,$\mu$m) in terms of atmospheric scale height. A negative number here indicates that the blue-optical is lower in altitude than the mid-IR, with the opposite true for positive numbers. We show the model trends for the uniform cloud, scattering haze, and super solar metallicity parameters covered by both grids (Fig. \ref{fig:spectral_index}). 

The transmission spectral indices show that the differences between local condensation and rainout condensation are predominantly in the $\Delta z_{UB-LM}/H_{eq}$ index, where $\Delta z_{UB-LM}/H_{eq}$\,=\,-0.8 and -1.75 for the clear solar model for local condensation and rainout condensation respectively. In all model cases the H$_2$O amplitude is decreased, with the exception of the 10$\times$ solar model which would require the optical and mid-IR data to distinguish from a clear solar case. There is a distinct separation between each of the model tracks. Increasing the metallicity decreases the amplitude of the water feature while increasing the radius ratio in the mid-IR compared to the optical, mainly due to the presence of CO$_2$ at higher metallicities (see Fig. \ref{fig:ggg_rainout_test}). Increasing the uniform cloud parameter decreases the H$_2$O amplitude with little change to the relative radius ratio between the optical and mid-IR. Increasing the wavelength dependent scattering haze decreases the amplitude of the H$_2$O feature while increasing the relative radius ratio between the optical and mid-IR. 

We note that these computed transmission spectral tracks show the trend of the individual parameters compared to a clear solar abundance atmosphere and for example, do not depict the combined impact of scattering haze and high metallicity. However, these are a useful indication of dominant factors in the transmission spectra of exoplanets and have successfully been used to make predictions for atmospheric measurements \citep{Sing2016,Wakeford2018}. Each of the measurements shown on the diagram are from \citet{Sing2016} with the addition of WASP-39b from \citet{Wakeford2018}. A majority of the currently measured transmission spectra from the optical to the mid-IR follow the track of increased scattering haze with a majority of measured H$_2$O amplitudes between 20--50\% of the clear solar model. These indices demonstrate the importance of the optical and mid-IR data to distinguish between different model parameters, in this case the cloud, haze, and enhanced metallicity trends. 

\section{Conclusions}
\label{sec:conclusions}
We present a publicly available\footnote{\url{https://drive.google.com/open?id=1ZFbkPdqg37_Om7ECSspSpEp5QrUMfA9J}} generic forward model grid of exoplanet transmission spectra, computed using the \texttt{ATMO} code, based on both local condensation and rainout condensation conditions. This grid can be scaled to a wide range of H$_2$/He dominated planetary atmospheres. The entire grid consists of \corr{56,320} model simulations across 22 isothermal temperatures, four planetary gravities, five atmospheric metallicities, \corr{four C/O ratios}, four uniform cloud parameters, four scattering haze parameters, and two chemical condensation scenarios. We demonstrate by sensitivity analysis, the reasoning behind selection of different grid parameters and their values. We derive scaling equations which can be used with this grid, for a wide range of planet-star combination. This grid of forward models is validated against published models of \citet{Fortney2010}, \citet{Kempton2017} and \citet{Goyal2018}.

\corr{Degeneracies are known to exist in the interpretation of various characteristics of planetary atmospheres from observations. The grid of atmospheric models presented here allows us to decouple and better understand the thermochemical processes shaping observable spectra. We demonstrate how changes in temperature and chemical abundances cause different trends in the spectra, and identify the following major factors that can effect interpretation of observations.} 
\begin{itemize}
\item \corr{Adopting different condensation processes (rainout and local) can lead to different interpretation of observations.}
\item \corr{SO$_2$ features at 6--8\textmu m, along with H$_2$O, can be used to constrain the metallicity of the exoplanet atmosphere, since the SO$_2$ spectral feature only appears for metallcities greater than 100x solar.}
\item \corr{The presence of VO without TiO can help constrain the temperature of the atmospheric limb, and that both TiO/VO features can reveal dominant physical process (rainout or local condensation) in the planet's atmosphere.}
\item \corr{At high C/O ratios {($\sim$1)}, spectral features in the infrared are different between the rainout and local condensation case, because rainout case has higher abundance of carbon bearing species without oxygen such as CH$_4$, C$_2$H$_2$ and HCN.}
\item \corr{The difference in solar elemental abundances between \citet{Asplund2009} and \citet{Lodders2003} used for model initialisation, can lead to differences in equilibrium chemical abundances and therefore the spectral features.}
\end{itemize}

This grid can be used to interpret the observations of H$_2$/He dominated hot jupiter exoplanet atmospheres, as well as to plan future observations using the HST, VLT, JWST and various other telescopes. \corr{It can be used directly with the HST and JWST simulator \texttt{PandExo} \citep{Batalha2017} for planning observations. The scaling flexibility provided by this grid for a wide range of planet-star combinations will be extremely valuable to efficiently choose and plan observations, for soon to be discovered TESS targets.}The fine variation of parameters in the grid also allows it to be incorporated in a retrieval framework with various machine learning techniques, as demonstrated in \citet{machinelearning2018}.

\section*{Acknowledgements}
\corr{We would like to thank the anonymous reviewer for their constructive comments that improved the paper substantially.} We would like to thank Mark Marley and Channon Visscher for valuable discussions on rainout condensation and comparative studies to their chemical models.
J.M.G and N.M are part funded by a Leverhulme Trust Research Project Grant, and in part by a University of Exeter College of Engineering, Mathematics and Physical Sciences PhD studentship.
H.R.W acknowledges funding by Association of Universities for Research in Astronomy (AURA) through a Giacconi Fellowship appointed at Space Telescope Science Institute (STScI). D.K.S and B.D acknowledge support from the European Research Council under the European Unions Seventh Framework Programme (FP7/2007-2013)/ ERC grant agreement number 336792. We acknowledge support from the Space Telescope Science Institute (STScI) Research Visitors Program.

This work used the DiRAC Complexity system, operated by the University of Leicester IT Services, which forms part of the STFC DiRAC HPC Facility. This work also used the University of Exeter Supercomputer, a DiRAC Facility jointly funded by STFC, the Large Facilities Capital Fund of BIS and the University of Exeter.

\subsection*{Author Contributions}
J.M.G computed the model grids. J.M.G, B.D, and H.R.W benchmarked the models with the help of N.K.L, D.K.S and N.M. J.M.G and H.R.W wrote the manuscript. H.R.W made the figures. All authors commented on and approved the final manuscript.

\bibliographystyle{mnras}
\bibliography{ggg_bibliography} 

\begin{thebibliography}{}
\makeatletter
\relax
\def\mn@urlcharsother{\let\do\@makeother \do\$\do\&\do\#\do\^\do\_\do\%\do\~}
\def\mn@doi{\begingroup\mn@urlcharsother \@ifnextchar [ {\mn@doi@}
  {\mn@doi@[]}}
\def\mn@doi@[#1]#2{\def\@tempa{#1}\ifx\@tempa\@empty \href
  {http://dx.doi.org/#2} {doi:#2}\else \href {http://dx.doi.org/#2} {#1}\fi
  \endgroup}
\def\mn@eprint#1#2{\mn@eprint@#1:#2::\@nil}
\def\mn@eprint@arXiv#1{\href {http://arxiv.org/abs/#1} {{\tt arXiv:#1}}}
\def\mn@eprint@dblp#1{\href {http://dblp.uni-trier.de/rec/bibtex/#1.xml}
  {dblp:#1}}
\def\mn@eprint@#1:#2:#3:#4\@nil{\def\@tempa {#1}\def\@tempb {#2}\def\@tempc
  {#3}\ifx \@tempc \@empty \let \@tempc \@tempb \let \@tempb \@tempa \fi \ifx
  \@tempb \@empty \def\@tempb {arXiv}\fi \@ifundefined
  {mn@eprint@\@tempb}{\@tempb:\@tempc}{\expandafter \expandafter \csname
  mn@eprint@\@tempb\endcsname \expandafter{\@tempc}}}

\bibitem[\protect\citeauthoryear{{Allard}, {Allard}, {Hauschildt}, {Kielkopf}
  \& {Machin}}{{Allard} et~al.}{2003}]{Allard2003}
{Allard} N.~F.,  {Allard} F.,  {Hauschildt} P.~H.,  {Kielkopf} J.~F.,
  {Machin} L.,  2003, \mn@doi [\aap] {10.1051/0004-6361:20031299}, \href
  {http://adsabs.harvard.edu/abs/2003A%26A...411L.473A} {411, L473}

\bibitem[\protect\citeauthoryear{{Amundsen}}{{Amundsen}}{2015}]{Amundsenthesis}
{Amundsen} D.~S.,  2015, PhD thesis

\bibitem[\protect\citeauthoryear{{Amundsen}, {Baraffe}, {Tremblin}, {Manners},
  {Hayek}, {Mayne}  \& {Acreman}}{{Amundsen} et~al.}{2014}]{Amundsen2014}
{Amundsen} D.~S.,  {Baraffe} I.,  {Tremblin} P.,  {Manners} J.,  {Hayek} W.,
  {Mayne} N.~J.,   {Acreman} D.~M.,  2014, \mn@doi [\aap]
  {10.1051/0004-6361/201323169}, \href
  {http://adsabs.harvard.edu/abs/2014A%26A...564A..59A} {564, A59}

\bibitem[\protect\citeauthoryear{{Asplund}, {Grevesse}, {Sauval}  \&
  {Scott}}{{Asplund} et~al.}{2009}]{Asplund2009}
{Asplund} M.,  {Grevesse} N.,  {Sauval} A.~J.,   {Scott} P.,  2009, \mn@doi
  [\araa] {10.1146/annurev.astro.46.060407.145222}, \href
  {http://adsabs.harvard.edu/abs/2009ARA%26A..47..481A} {47, 481}

\bibitem[\protect\citeauthoryear{{Batalha} et~al.,}{{Batalha}
  et~al.}{2017}]{Batalha2017}
{Batalha} N.~E.,  et~al., 2017, preprint, \href
  {http://adsabs.harvard.edu/abs/2017arXiv170201820B} {} (\mn@eprint {arXiv}
  {1702.01820})

\bibitem[\protect\citeauthoryear{{Burrows} \& {Sharp}}{{Burrows} \&
  {Sharp}}{1999}]{Burrows1999}
{Burrows} A.,  {Sharp} C.~M.,  1999, \mn@doi [\apj] {10.1086/306811}, \href
  {http://adsabs.harvard.edu/abs/1999ApJ...512..843B} {512, 843}

\bibitem[\protect\citeauthoryear{{Burrows}, {Marley}  \& {Sharp}}{{Burrows}
  et~al.}{2000}]{Burrows2000}
{Burrows} A.,  {Marley} M.~S.,   {Sharp} C.~M.,  2000, \mn@doi [\apj]
  {10.1086/308462}, \href
  {https://ui.adsabs.harvard.edu/#abs/2000ApJ...531..438B} {531, 438}

\bibitem[\protect\citeauthoryear{{Deming} et~al.,}{{Deming}
  et~al.}{2013}]{Deming2013}
{Deming} D.,  et~al., 2013, \mn@doi [\apj] {10.1088/0004-637X/774/2/95}, \href
  {http://adsabs.harvard.edu/abs/2013ApJ...774...95D} {774, 95}

\bibitem[\protect\citeauthoryear{{Drummond}}{{Drummond}}{2017}]{drummond2017}
{Drummond} B.,  2017, PhD thesis, University of Exeter

\bibitem[\protect\citeauthoryear{{Drummond}, {Tremblin}, {Baraffe}, {Amundsen},
  {Mayne}, {Venot}  \& {Goyal}}{{Drummond} et~al.}{2016}]{Drummond2016}
{Drummond} B.,  {Tremblin} P.,  {Baraffe} I.,  {Amundsen} D.~S.,  {Mayne}
  N.~J.,  {Venot} O.,   {Goyal} J.,  2016, \mn@doi [\aap]
  {10.1051/0004-6361/201628799}, \href
  {http://adsabs.harvard.edu/abs/2016A%26A...594A..69D} {594, A69}

\bibitem[\protect\citeauthoryear{{Evans} et~al.,}{{Evans}
  et~al.}{2016}]{Evans2016}
{Evans} T.~M.,  et~al., 2016, \mn@doi [\apjl] {10.3847/2041-8205/822/1/L4},
  \href {http://adsabs.harvard.edu/abs/2016ApJ...822L...4E} {822, L4}

\bibitem[\protect\citeauthoryear{{Evans} et~al.,}{{Evans}
  et~al.}{2017}]{Evans2017}
{Evans} T.~M.,  et~al., 2017, \mn@doi [\nat] {10.1038/nature23266}, \href
  {http://adsabs.harvard.edu/abs/2017Natur.548...58E} {548, 58}

\bibitem[\protect\citeauthoryear{{Fortney}}{{Fortney}}{2005}]{Fortney2005}
{Fortney} J.~J.,  2005, \mn@doi [\mnras] {10.1111/j.1365-2966.2005.09587.x},
  \href {http://adsabs.harvard.edu/abs/2005MNRAS.364..649F} {364, 649}

\bibitem[\protect\citeauthoryear{{Fortney}, {Lodders}, {Marley}  \&
  {Freedman}}{{Fortney} et~al.}{2008}]{Fortney2008}
{Fortney} J.~J.,  {Lodders} K.,  {Marley} M.~S.,   {Freedman} R.~S.,  2008,
  \mn@doi [\apj] {10.1086/528370}, \href
  {http://adsabs.harvard.edu/abs/2008ApJ...678.1419F} {678, 1419}

\bibitem[\protect\citeauthoryear{{Fortney}, {Shabram}, {Showman}, {Lian},
  {Freedman}, {Marley}  \& {Lewis}}{{Fortney} et~al.}{2010}]{Fortney2010}
{Fortney} J.~J.,  {Shabram} M.,  {Showman} A.~P.,  {Lian} Y.,  {Freedman}
  R.~S.,  {Marley} M.~S.,   {Lewis} N.~K.,  2010, \mn@doi [\apj]
  {10.1088/0004-637X/709/2/1396}, \href
  {http://adsabs.harvard.edu/abs/2010ApJ...709.1396F} {709, 1396}

\bibitem[\protect\citeauthoryear{{Fortney} et~al.,}{{Fortney}
  et~al.}{2016}]{Fortney2016}
{Fortney} J.~J.,  et~al., 2016, preprint, \href
  {http://adsabs.harvard.edu/abs/2016arXiv160206305F} {} (\mn@eprint {arXiv}
  {1602.06305})

\bibitem[\protect\citeauthoryear{{Freedman}, {Marley}  \& {Lodders}}{{Freedman}
  et~al.}{2008}]{Freedman2008}
{Freedman} R.~S.,  {Marley} M.~S.,   {Lodders} K.,  2008, \mn@doi [\apjs]
  {10.1086/521793}, \href {http://adsabs.harvard.edu/abs/2008ApJS..174..504F}
  {174, 504}

\bibitem[\protect\citeauthoryear{{Goyal} et~al.,}{{Goyal}
  et~al.}{2018}]{Goyal2018}
{Goyal} J.~M.,  et~al., 2018, \mn@doi [\mnras] {10.1093/mnras/stx3015}, \href
  {https://ui.adsabs.harvard.edu/#abs/2018MNRAS.474.5158G} {474, 5158}

\bibitem[\protect\citeauthoryear{{Heng} \& {Kitzmann}}{{Heng} \&
  {Kitzmann}}{2017}]{Heng2017}
{Heng} K.,  {Kitzmann} D.,  2017, preprint, \href
  {http://adsabs.harvard.edu/abs/2017arXiv170202051H} {} (\mn@eprint {arXiv}
  {1702.02051})

\bibitem[\protect\citeauthoryear{{Kempton}, {Lupu}, {Owusu-Asare}, {Slough}  \&
  {Cale}}{{Kempton} et~al.}{2017}]{Kempton2017}
{Kempton} E.~M.-R.,  {Lupu} R.,  {Owusu-Asare} A.,  {Slough} P.,   {Cale} B.,
  2017, \mn@doi [\pasp] {10.1088/1538-3873/aa61ef}, \href
  {http://adsabs.harvard.edu/abs/2017PASP..129d4402K} {129, 044402}

\bibitem[\protect\citeauthoryear{{Kreidberg} et~al.,}{{Kreidberg}
  et~al.}{2014}]{Kreidberg2014}
{Kreidberg} L.,  et~al., 2014, \mn@doi [\apjl] {10.1088/2041-8205/793/2/L27},
  \href {http://adsabs.harvard.edu/abs/2014ApJ...793L..27K} {793, L27}

\bibitem[\protect\citeauthoryear{{Lecavelier Des Etangs}, {Pont},
  {Vidal-Madjar}  \& {Sing}}{{Lecavelier Des Etangs}
  et~al.}{2008}]{Lecavelier2008}
{Lecavelier Des Etangs} A.,  {Pont} F.,  {Vidal-Madjar} A.,   {Sing} D.,  2008,
  \mn@doi [\aap] {10.1051/0004-6361:200809388}, \href
  {http://adsabs.harvard.edu/abs/2008A%26A...481L..83L} {481, L83}

\bibitem[\protect\citeauthoryear{{Lodders}}{{Lodders}}{1999}]{Lodders1999}
{Lodders} K.,  1999, \mn@doi [\apj] {10.1086/307387}, \href
  {http://adsabs.harvard.edu/abs/1999ApJ...519..793L} {519, 793}

\bibitem[\protect\citeauthoryear{{Lodders}}{{Lodders}}{2002}]{Lodders2002}
{Lodders} K.,  2002, \mn@doi [\apj] {10.1086/342241}, \href
  {http://adsabs.harvard.edu/abs/2002ApJ...577..974L} {577, 974}

\bibitem[\protect\citeauthoryear{{Lodders}}{{Lodders}}{2003}]{Lodders2003}
{Lodders} K.,  2003, \mn@doi [\apj] {10.1086/375492}, \href
  {https://ui.adsabs.harvard.edu/#abs/2003ApJ...591.1220L} {591, 1220}

\bibitem[\protect\citeauthoryear{{Lodders}}{{Lodders}}{2006}]{Lodders2006}
{Lodders} K.,  2006, \mn@doi [\apjl] {10.1086/507181}, \href
  {http://adsabs.harvard.edu/abs/2006ApJ...647L..37L} {647, L37}

\bibitem[\protect\citeauthoryear{{Lodders} \& {Fegley}}{{Lodders} \&
  {Fegley}}{2002}]{LoddersFegley2002}
{Lodders} K.,  {Fegley} B.,  2002, \mn@doi [\icarus] {10.1006/icar.2001.6740},
  \href {http://adsabs.harvard.edu/abs/2002Icar..155..393L} {155, 393}

\bibitem[\protect\citeauthoryear{{Lodders} \& {Fegley}}{{Lodders} \&
  {Fegley}}{2006}]{LoddersFegley2006}
{Lodders} K.,  {Fegley} Jr. B.,  2006, {Chemistry of Low Mass Substellar
  Objects}.
p.~1, \mn@doi{10.1007/3-540-30313-8_1}

\bibitem[\protect\citeauthoryear{{Madhusudhan}}{{Madhusudhan}}{2012}]{Madhusudhan2012}
{Madhusudhan} N.,  2012, \mn@doi [\apj] {10.1088/0004-637X/758/1/36}, \href
  {http://adsabs.harvard.edu/abs/2012ApJ...758...36M} {758, 36}

\bibitem[\protect\citeauthoryear{{Marquez-Neila}, {Fisher}, {Sznitman}  \&
  {Heng}}{{Marquez-Neila} et~al.}{2018}]{machinelearning2018}
{Marquez-Neila} P.,  {Fisher} C.,  {Sznitman} R.,   {Heng} K.,  2018, preprint,
  \href {https://ui.adsabs.harvard.edu/#abs/2018arXiv180603944M} {p.
  arXiv:1806.03944} (\mn@eprint {arXiv} {1806.03944})

\bibitem[\protect\citeauthoryear{{Mbarek} \& {Kempton}}{{Mbarek} \&
  {Kempton}}{2016}]{Mbarek2016}
{Mbarek} R.,  {Kempton} E.~M.-R.,  2016, \mn@doi [\apj]
  {10.3847/0004-637X/827/2/121}, \href
  {http://adsabs.harvard.edu/abs/2016ApJ...827..121M} {827, 121}

\bibitem[\protect\citeauthoryear{{Molli{\`e}re}, {van Boekel}, {Dullemond},
  {Henning}  \& {Mordasini}}{{Molli{\`e}re} et~al.}{2015}]{Molliere2015}
{Molli{\`e}re} P.,  {van Boekel} R.,  {Dullemond} C.,  {Henning} T.,
  {Mordasini} C.,  2015, \mn@doi [\apj] {10.1088/0004-637X/813/1/47}, \href
  {http://adsabs.harvard.edu/abs/2015ApJ...813...47M} {813, 47}

\bibitem[\protect\citeauthoryear{{Molli{\`e}re}, {van Boekel}, {Bouwman},
  {Henning}, {Lagage}  \& {Min}}{{Molli{\`e}re} et~al.}{2016}]{Molliere2016}
{Molli{\`e}re} P.,  {van Boekel} R.,  {Bouwman} J.,  {Henning} T.,  {Lagage}
  P.-O.,   {Min} M.,  2016, preprint, \href
  {http://adsabs.harvard.edu/abs/2016arXiv161108608M} {} (\mn@eprint {arXiv}
  {1611.08608})

\bibitem[\protect\citeauthoryear{{Moses} et~al.,}{{Moses}
  et~al.}{2013}]{Moses2013b}
{Moses} J.~I.,  et~al., 2013, \mn@doi [\apj] {10.1088/0004-637X/777/1/34},
  \href {http://adsabs.harvard.edu/abs/2013ApJ...777...34M} {777, 34}

\bibitem[\protect\citeauthoryear{{Nikolov} et~al.,}{{Nikolov}
  et~al.}{2018}]{Nikolov2018}
{Nikolov} N.,  et~al., 2018, \mn@doi [\nat] {10.1038/s41586-018-0101-7}, \href
  {https://ui.adsabs.harvard.edu/#abs/2018Natur.557..526N} {557, 526}

\bibitem[\protect\citeauthoryear{{Rothman} et~al.,}{{Rothman}
  et~al.}{2010}]{Rothman2010}
{Rothman} L.~S.,  et~al., 2010, \mn@doi [\jqsrt] {10.1016/j.jqsrt.2010.05.001},
  \href {http://adsabs.harvard.edu/abs/2010JQSRT.111.2139R} {111, 2139}

\bibitem[\protect\citeauthoryear{{Rothman} et~al.,}{{Rothman}
  et~al.}{2013}]{Rothman2013}
{Rothman} L.~S.,  et~al., 2013, \mn@doi [\jqsrt] {10.1016/j.jqsrt.2013.07.002},
  \href {http://adsabs.harvard.edu/abs/2013JQSRT.130....4R} {130, 4}

\bibitem[\protect\citeauthoryear{{Sing} et~al.,}{{Sing}
  et~al.}{2016}]{Sing2016}
{Sing} D.~K.,  et~al., 2016, \mn@doi [\nat] {10.1038/nature16068}, \href
  {http://adsabs.harvard.edu/abs/2016Natur.529...59S} {529, 59}

\bibitem[\protect\citeauthoryear{{Southworth}}{{Southworth}}{2011}]{Southworth:2011aa}
{Southworth} J.,  2011, \mn@doi [\mnras] {10.1111/j.1365-2966.2011.19399.x},
  \href {http://adsabs.harvard.edu/abs/2011MNRAS.417.2166S} {417, 2166}

\bibitem[\protect\citeauthoryear{{Tennyson} et~al.,}{{Tennyson}
  et~al.}{2016}]{Tennyson2016}
{Tennyson} J.,  et~al., 2016, \mn@doi [Journal of Molecular Spectroscopy]
  {10.1016/j.jms.2016.05.002}, \href
  {http://adsabs.harvard.edu/abs/2016JMoSp.327...73T} {327, 73}

\bibitem[\protect\citeauthoryear{{Tremblin}, {Amundsen}, {Mourier}, {Baraffe},
  {Chabrier}, {Drummond}, {Homeier}  \& {Venot}}{{Tremblin}
  et~al.}{2015}]{Tremblin2015}
{Tremblin} P.,  {Amundsen} D.~S.,  {Mourier} P.,  {Baraffe} I.,  {Chabrier} G.,
   {Drummond} B.,  {Homeier} D.,   {Venot} O.,  2015, \mn@doi [\apjl]
  {10.1088/2041-8205/804/1/L17}, \href
  {http://adsabs.harvard.edu/abs/2015ApJ...804L..17T} {804, L17}

\bibitem[\protect\citeauthoryear{{Tremblin}, {Amundsen}, {Chabrier}, {Baraffe},
  {Drummond}, {Hinkley}, {Mourier}  \& {Venot}}{{Tremblin}
  et~al.}{2016}]{Tremblin2016}
{Tremblin} P.,  {Amundsen} D.~S.,  {Chabrier} G.,  {Baraffe} I.,  {Drummond}
  B.,  {Hinkley} S.,  {Mourier} P.,   {Venot} O.,  2016, \mn@doi [\apjl]
  {10.3847/2041-8205/817/2/L19}, \href
  {http://adsabs.harvard.edu/abs/2016ApJ...817L..19T} {817, L19}

\bibitem[\protect\citeauthoryear{{Tremblin} et~al.,}{{Tremblin}
  et~al.}{2017}]{Tremblin2017}
{Tremblin} P.,  et~al., 2017, \mn@doi [\apj] {10.3847/1538-4357/aa6e57}, \href
  {https://ui.adsabs.harvard.edu/#abs/2017ApJ...841...30T} {841, 30}

\bibitem[\protect\citeauthoryear{{Vandaele} et~al.,}{{Vandaele}
  et~al.}{2017}]{Vandaele2017}
{Vandaele} A.~C.,  et~al., 2017, \mn@doi [\icarus]
  {10.1016/j.icarus.2017.05.001}, \href
  {https://ui.adsabs.harvard.edu/#abs/2017Icar..295....1V} {295, 1}

\bibitem[\protect\citeauthoryear{{Visscher}, {Lodders}  \& {Fegley}}{{Visscher}
  et~al.}{2006}]{Visscher2006}
{Visscher} C.,  {Lodders} K.,   {Fegley} Jr. B.,  2006, \mn@doi [\apj]
  {10.1086/506245}, \href {http://adsabs.harvard.edu/abs/2006ApJ...648.1181V}
  {648, 1181}

\bibitem[\protect\citeauthoryear{{Wakeford}, {Sing}, {Evans}, {Deming}  \&
  {Mandell}}{{Wakeford} et~al.}{2016}]{Wakeford2016}
{Wakeford} H.~R.,  {Sing} D.~K.,  {Evans} T.,  {Deming} D.,   {Mandell} A.,
  2016, \mn@doi [\apj] {10.3847/0004-637X/819/1/10}, \href
  {http://adsabs.harvard.edu/abs/2016ApJ...819...10W} {819, 10}

\bibitem[\protect\citeauthoryear{{Wakeford} et~al.,}{{Wakeford}
  et~al.}{2018}]{Wakeford2018}
{Wakeford} H.~R.,  et~al., 2018, \mn@doi [\aj] {10.3847/1538-3881/aa9e4e},
  \href {https://ui.adsabs.harvard.edu/#abs/2018AJ....155...29W} {155}

\makeatother
\end{thebibliography}








\bsp	
\label{lastpage}
\end{document}